\pdfoutput=1

\documentclass[11pt,twoside,a4paper,cmspaper,final,collab]{cms-tdr}
\def\svnVersion{242671}\def\cmsCernNo{2013-217}\def\cmsCernDate{\today}\def\cmsMessage{Published in Physical Review Letters as \href{http://dx.doi.org/10.1103/PhysRevLett.112.191802}{\doi{10.1103/PhysRevLett.112.191802.}}}

\begin{document}\cmsNoteHeader{SMP-12-011}

\hyphenation{had-ron-i-za-tion}
\hyphenation{cal-or-i-me-ter}
\hyphenation{de-vices}

\RCS$Revision: 235407 $
\RCS$HeadURL: svn+ssh://svn.cern.ch/reps/tdr2/papers/SMP-12-011/trunk/SMP-12-011.tex $
\RCS$Id: SMP-12-011.tex 235407 2014-04-04 16:53:44Z alverson $
\newlength\cmsFigWidth
\ifthenelse{\boolean{cms@external}}{\setlength\cmsFigWidth{0.85\columnwidth}}{\setlength\cmsFigWidth{0.4\textwidth}}
\ifthenelse{\boolean{cms@external}}{\providecommand{\cmsLeft}{top}}{\providecommand{\cmsLeft}{left}}
\ifthenelse{\boolean{cms@external}}{\providecommand{\cmsRight}{bottom}}{\providecommand{\cmsRight}{right}}
\ifthenelse{\boolean{cms@external}}
{\providecommand{\suppMaterial}{the Supplemental Material~\cite{suppRef}\xspace}}
{\providecommand{\suppMaterial}{Appendix~\ref{sec:SuppMaterial}\xspace}}
\ifthenelse{\boolean{cms@external}}{\providecommand{\NA}{\ensuremath{\cdots}}}{\providecommand{\NA}{\text{---}}}
\ifthenelse{\boolean{cms@external}}{\providecommand{\CL}{\ensuremath{\mathrm{C.L.}}\xspace}}{\providecommand{\CL}{\ensuremath{\mathrm{CL}}\xspace}}
\newcommand\THELUMI{\ensuremath{18.2 \pm 0.5\pbinv}\xspace}
\newcommand\THEORYSIGBRWP{\ensuremath{7.12 \pm 0.20\unit{nb}}\xspace}
\newcommand\THEORYSIGBRWM{\ensuremath{5.06 \pm 0.13\unit{nb}}\xspace}
\newcommand\THEORYSIGBRZ{\ensuremath{1.13 \pm 0.04\unit{nb}}\xspace}
\cmsNoteHeader{SMP-12-011} 
\title{\texorpdfstring{Measurement of inclusive $\PW$ and $\cPZ$ boson production cross sections in $\Pp\Pp$
  collisions at $\sqrt{s} =8\TeV$}{Measurement of inclusive W and Z boson production cross sections in pp collisions at sqrt(s)=8 TeV}}
\date{\today}

\abstract{
A measurement of total and fiducial inclusive $\PW$ and $\cPZ$ boson production cross sections in $\Pp\Pp$ collisions at $\sqrt{s}=8$\TeV is presented. Electron and muon final
states are analyzed in a data sample collected with the CMS
detector corresponding to an integrated luminosity of $18.2 \pm 0.5\pbinv$.
The measured total inclusive cross sections times branching fractions are
$\sigma(\Pp\Pp \to \PW X) \times \mathcal{B}(\PW \to \ell\nu ) =
12.21 \pm 0.03\stat \pm 0.24\syst\pm
0.32\lum\unit{nb}$ and $\sigma(\Pp\Pp \to \cPZ X) \times \mathcal{B}(\cPZ \to \ell^+\ell^-) =
1.15 \pm 0.01\stat \pm 0.02\syst\pm
0.03\lum\unit{nb}$ for the dilepton mass in the range of 60
to 120\GeV. The measured values
agree with next-to-next-to-leading-order QCD cross section
calculations. Ratios of cross sections are reported with a
precision of 2\%. This is the first measurement of inclusive $\PW$ and $\cPZ$ boson production in proton-proton collisions at $\sqrt{s}=8$\TeV.}

\hypersetup{%
pdfauthor={CMS Collaboration},%
pdftitle={Measurement of inclusive W and Z boson production cross sections in pp collisions at sqrt(s)= 8 TeV},%
pdfsubject={CMS},%
pdfkeywords={CMS, physics, W boson, Z boson}}

\maketitle

The production of $\PW$ and $\cPZ$ bosons is one of
the most prominent examples of hard scattering processes at hadron
colliders~\cite{PhysRevLett.25.316}. Theoretical predictions are available at next-to-next-to-leading order
(NNLO)~\cite{PhysRevD.51.44, Hamberg1991343, vanNeerven199211, PhysRevLett.88.201801, PhysRevD.69.094008} in
perturbative quantum chromodynamics (QCD). The calculations are limited by uncertainties in
parton distribution functions (PDFs), missing higher-order QCD effects, and
electroweak (EW) radiative corrections, which are available at next-to-leading order
(NLO)~\cite{horace1,horace2, horace3, horace4}. Precise measurements of inclusive cross sections provide
tests of perturbative QCD and validate the theoretical predictions of
higher-order corrections. Additionally, accurate measurements can be used to constrain PDFs.

Inclusive $\PW$ and $\cPZ$ boson production cross sections and their ratios were previously measured by the ATLAS and
CMS Collaborations at the Large Hadron Collider
(LHC) in  proton-proton collisions at  $\sqrt{s} =
7\TeV$~\cite{CMS-EWK-WZ, CMS-EWK-TAU, ATLAS-EWK-WZ}.
This Letter describes the inclusive measurement at $\sqrt{s} =
8\TeV$, performed in the electron and muon
decay channels, with the CMS detector.  A data sample
collected in 2012 corresponding to an integrated luminosity of
$\THELUMI$ is used.

The levels of instantaneous luminosity reached by the LHC in 2012
present challenges for the precise measurement of the $\PW$ boson cross section
because of the degraded missing transverse momentum resolution
resulting from the large number of $\Pp\Pp$ interactions per bunch crossing (pileup).
A data sample with low pileup was collected in May~2012
by adjusting the beam separation during data taking.
An average of 4 interactions per bunch crossing was achieved,
compared with the average of 21 during the rest of 2012.
The measurements of the $\PW$ and $\cPZ$ boson production cross sections are
performed using this data sample.

The central feature of the CMS apparatus is a superconducting
solenoid, of 6\unit{m} internal diameter, providing a field of 3.8\unit{T}. Within
the field volume are a silicon pixel and strip tracker, a crystal
electromagnetic calorimeter (ECAL), and a brass/scintillator hadron
calorimeter. Muons are measured in gas-ionization detectors embedded
in the steel flux-return yoke of the magnet.
CMS uses a right-handed coordinate system, with the origin at the
nominal interaction point, the $x$ axis pointing to the center of the
LHC, the $y$ axis pointing upwards, perpendicular to the plane of the
LHC ring, and the $z$ axis along the counterclockwise-beam
direction. The polar angle $\theta$ is measured from the positive $z$
axis, and the azimuthal angle $\phi$ is measured in the
$x$-$y$ plane. The pseudorapidity $\eta$ is
defined by $\eta = -\ln [\tan (\theta/2)]$.
Details of the CMS detector and its performance can be found elsewhere~\cite{CMS-JINST}.

Leptonic $\PW$ boson decays are characterized by a prompt, energetic, and
isolated charged lepton and a neutrino giving rise to significant missing
transverse energy, $\MET$. Events used in the cross section
measurement are not required to have a minimum reconstructed $\MET$, but
the $\MET$ distribution is used as a discriminant against background
from multijet events. $\cPZ$ boson decays to leptons are selected by requiring two energetic and
isolated leptons of the same flavor and opposite charge. The $\cPZ$ boson
candidates are required to have a reconstructed dilepton mass between
60 and 120\GeV. Samples of $\cPZ$ boson candidates satisfying looser lepton requirements are used to estimate
efficiencies.

Because of the high rate of collisions and the limited bandwidth for data processing,
the data acquisition system must be selective in deciding which events are
sufficiently interesting to be kept for analysis. Triggers make rapid
decisions by executing simplified muon and electron reconstruction algorithms.
For this analysis, the events are collected when triggered by the presence of a
muon with large transverse momentum, $\pt>15\GeV$, and $\abs{\eta}<2.1$,
or an electron with large transverse energy, $\ET>22$\GeV, and $\abs{\eta}<2.5$, with
loose isolation and identification requirements.

Electrons are identified as clusters of energy deposits in the ECAL
matched to tracks measured with the silicon
tracker~\cite{Adam:815410,CMS_tracking,Chatrchyan:2013dga,PhysRevLett.106.082001,elereco}.
The ECAL fiducial region is defined by $\abs{\eta}<1.44$
(barrel) or $1.57<\abs{\eta}<2.5$ (endcap), where
$\eta$ is the pseudorapidity of the energy cluster.  The
barrel-endcap transition region and the first ring of endcap trigger
towers are excluded because they are partially obscured by cables and
services  exiting between the barrel and endcaps. A cluster is
considered to be within the acceptance of the ECAL if it is within the
ECAL fiducial region and has transverse energy $\ET>25$\GeV.
Electrons are required to be isolated from other reconstructed particles in a cone of $\Delta R = 0.3$, where
$\Delta R = \sqrt{ \smash[b]{\Delta \eta^2 + \Delta \phi^2}}$.  Particle candidates are identified using a particle-flow
algorithm~\cite{CMS-PAS-PFT-10-001, CMS-PAS-PFT-10-003} that provides a complete description of the
event in terms of electrons, muons, photons, charged hadrons, and neutral hadrons.
An electron candidate is selected if the sum of transverse momenta of particles in the cone is less than 15\%
of the candidate's transverse energy.

Muons are reconstructed from seed tracks in the muon detector
combined with silicon strip and pixel information using a global fit~\cite{CMS-PAS-MUO-10-002, muonReco}.
In the \pt range of interest, the momentum resolution
is driven by the inner tracking system.
Muons with $\pt>25$\GeV and $\abs{\eta}<2.1$ are selected, which is
consistent with the acceptance of the single muon trigger.
A relative isolation variable is computed as discussed for electrons,
but in a cone of radius $\Delta R = 0.4$ and with an isolation selection
requirement of less than 12\%.

The acceptance for $\PW$ or $\cPZ$ boson events is the fraction of generated events for which the leptons satisfy the restrictions on $\eta$ and $\pt$. The event selection criteria will select a subset of the accepted events, and the efficiency
specifies the fraction of events selected. This accounts for example for the region of the ECAL from $1.44 < \abs{\eta} < 1.55$. Other effects, like crystal boundaries, are accounted in the efficiency to reconstruct leptons. Using this acceptance definition, we are able to separate experimental from theoretical uncertainties in the measurement. The detector response is simulated using a detailed description of the CMS detector, based on the \textsc{geant4} package~\cite{Agostinelli:2002hh}. Additional proton-proton interactions are taken into account using an admixture of simulated minimum bias events and the same reconstruction code is applied for
data and simulated events. Data to simulation ratios of efficiencies are used as scale factors. No single event generator gives a reliable description of both EW and QCD effects. The acceptance is estimated using Monte Carlo simulation based on \POWHEG~\cite{POWHEG-V, POWHEG1, POWHEG2, POWHEG3}. The effects of nonperturbative QCD, higher-order QCD, and electroweak corrections on the estimated acceptance are investigated using specific simulation tools, from which uncertainties are derived~\cite{resbos, photos, horace1, horace2, horace3, horace4}. The uncertainty related to the PDFs is estimated following closely the prescription of the PDF4LHC working group~\cite{PDF4LHC} to combine uncertainties related to the choice of the NLO PDF and the strong coupling constant, $\alpha_s$.

\begin{figure}[tbh]
\centering
        \includegraphics[width=.45\textwidth]{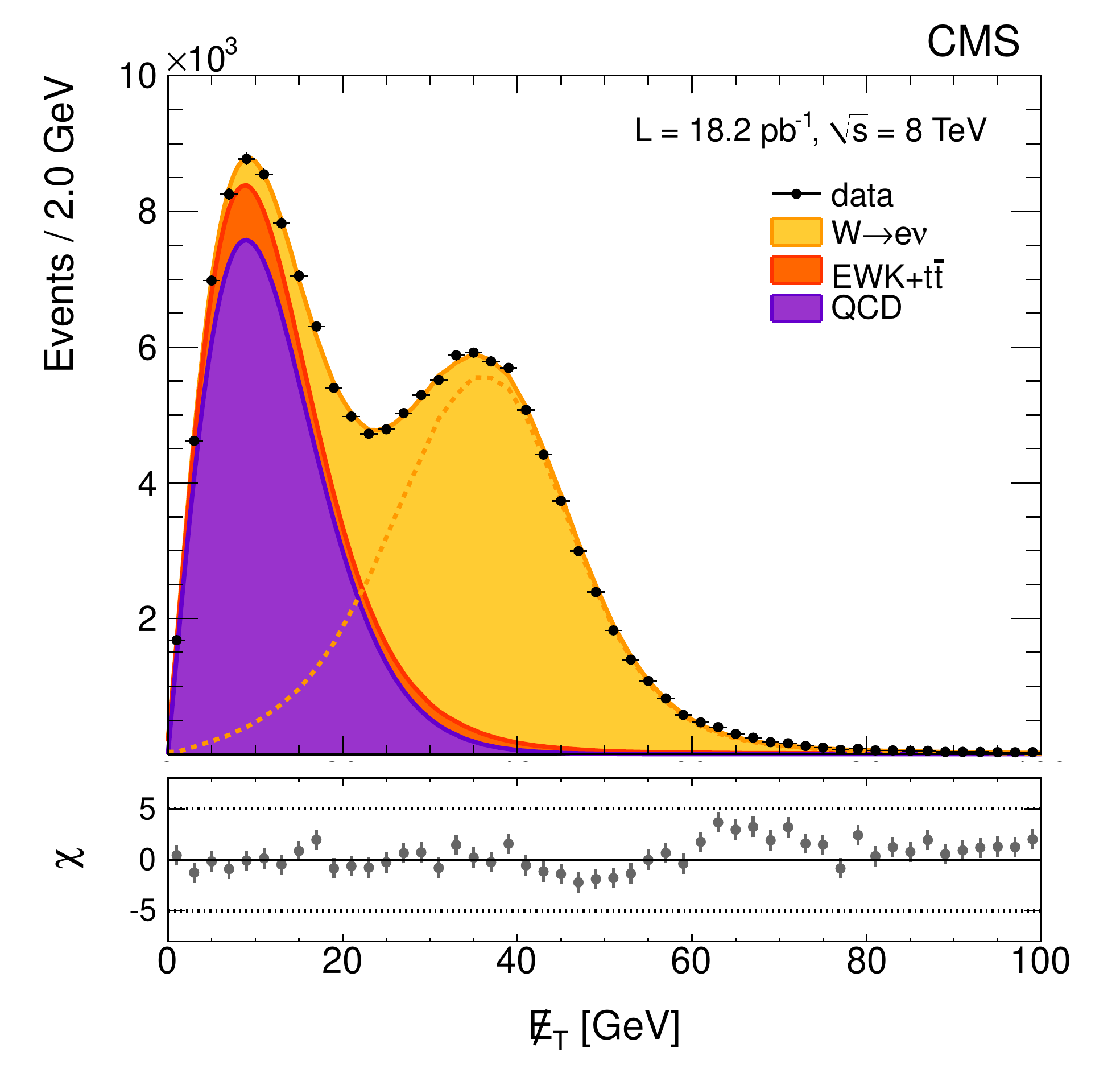}
         \includegraphics[width=.45\textwidth]{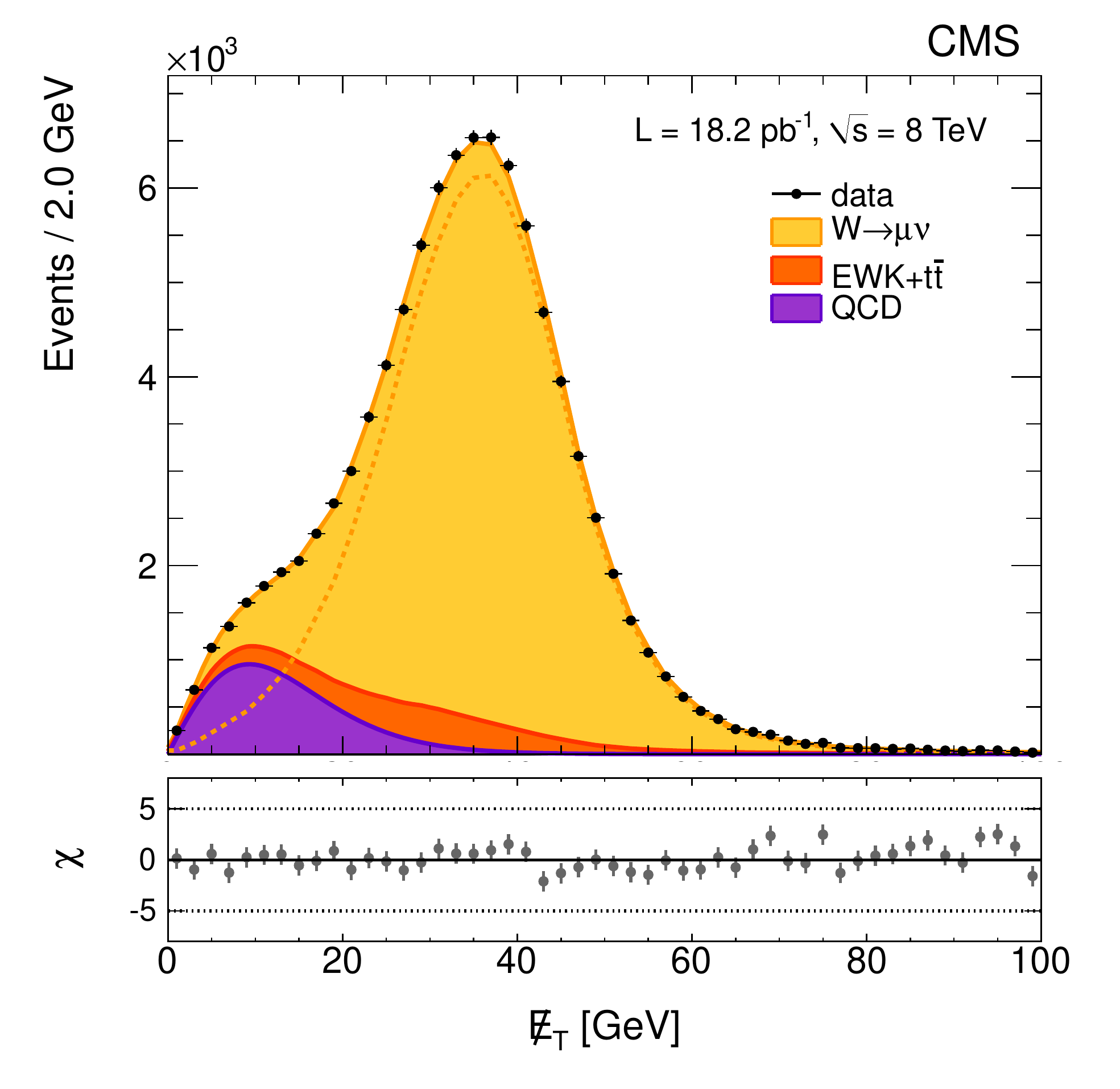}
       \caption{The missing transverse energy distributions for $\PW$
       boson candidate events in the electron (\cmsLeft)
       and muon (\cmsRight) final states. The variable $\chi$ shown in the
       lower plot is defined as
       $(N_{\text{obs}}-N_{\text{exp}})/\sqrt{N_{\text{obs}}}$,
       where $N_{\text{obs}}$ is the number of observed events and
       $N_{\text{exp}}$ is the total of the fitted signal and background yields.
       \label{fig:W}}
\end{figure}

The $\PW$ boson candidate events are required to have an identified
electron or muon. The $\PW$ boson signal and background yields are
obtained from the $\MET$ distributions using a binned
maximum-likelihood fit.
The missing transverse energy is calculated with
the particle-flow algorithm by adding the transverse energy vectors of all
identified particles.
An accurate $\MET$ measurement is essential to distinguish the $\PW$
signal from QCD multijet backgrounds.
To account for shortcomings of the simulation in modeling the recoil
against the $\PW$ boson, a correction is derived
from a $\cPZ$ boson sample. The recoil in these events is studied, in data
as well as in simulation, and the differences are propagated to the $\PW$ boson simulation as a function of the \pt of the generated $\PW$ or $\cPZ$ boson.
Other background processes from $\PW\to\tau\nu$, Drell--Yan, diboson, and top-pair production
also become significant at high $\MET$, contributing about 6\% of the total selected yield.
The background contribution from cosmic rays in the $\PW\to\Pgm\Pgn$ channel is negligible.
The $\MET$ model is fitted to the observed distribution as the sum of three contributions:
the $\PW$ signal, the QCD background, and other backgrounds. The QCD background is modeled by
an analytic function, while the signal and EW backgrounds are modeled with simulation-based
fitting functions~\cite{CMS-EWK-WZ}. The EW contributions are normalized to the $\PW$ signal yield in the fit
through the ratios of the theoretical cross sections. Figure~\ref{fig:W} shows the
$\MET$ distributions of the inclusive $\PW$ boson samples and the results of
the fit.

\begin{figure}[tbhp]
\centering
        \includegraphics[width=.45\textwidth]{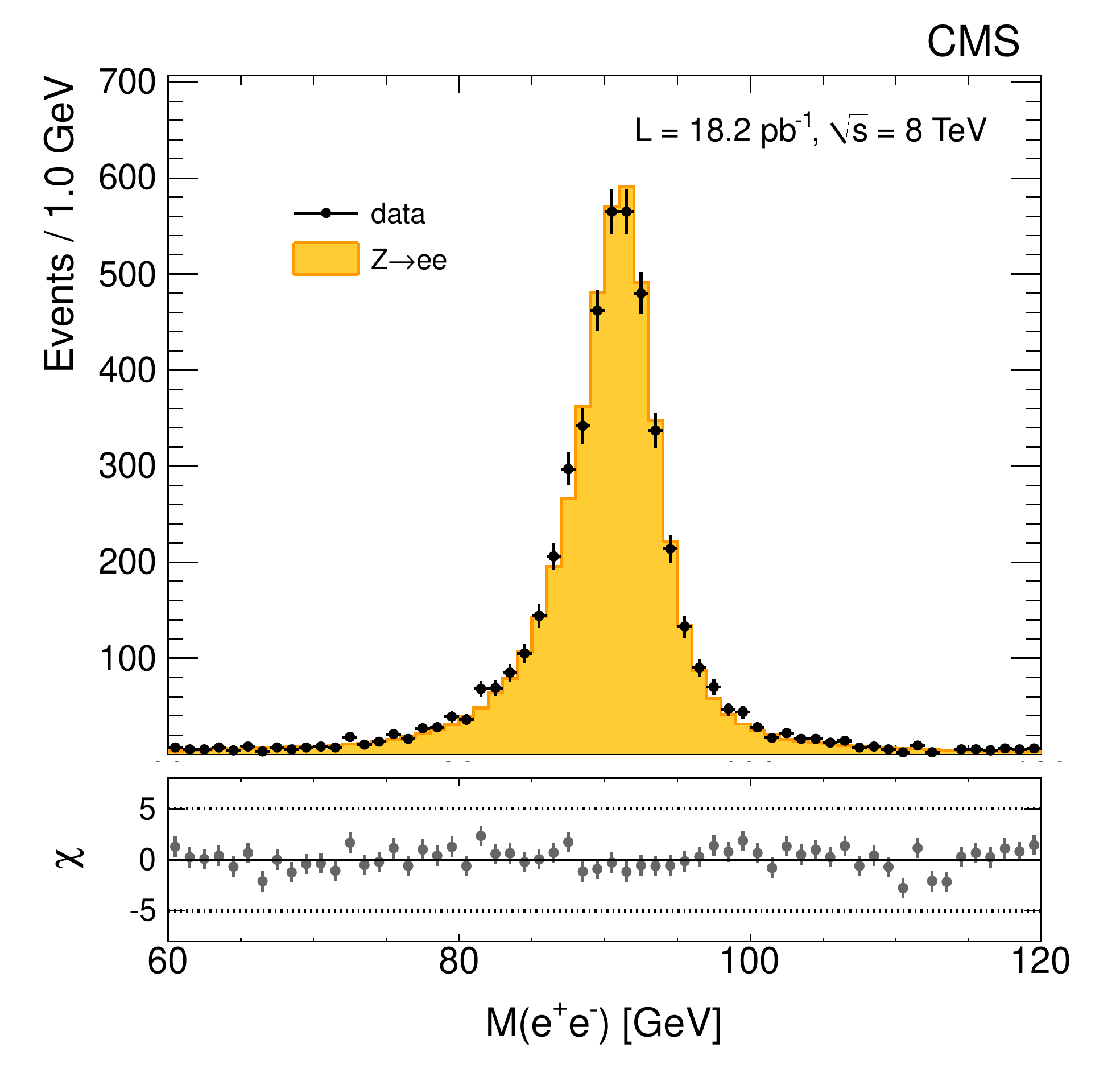}
         \includegraphics[width=.45\textwidth]{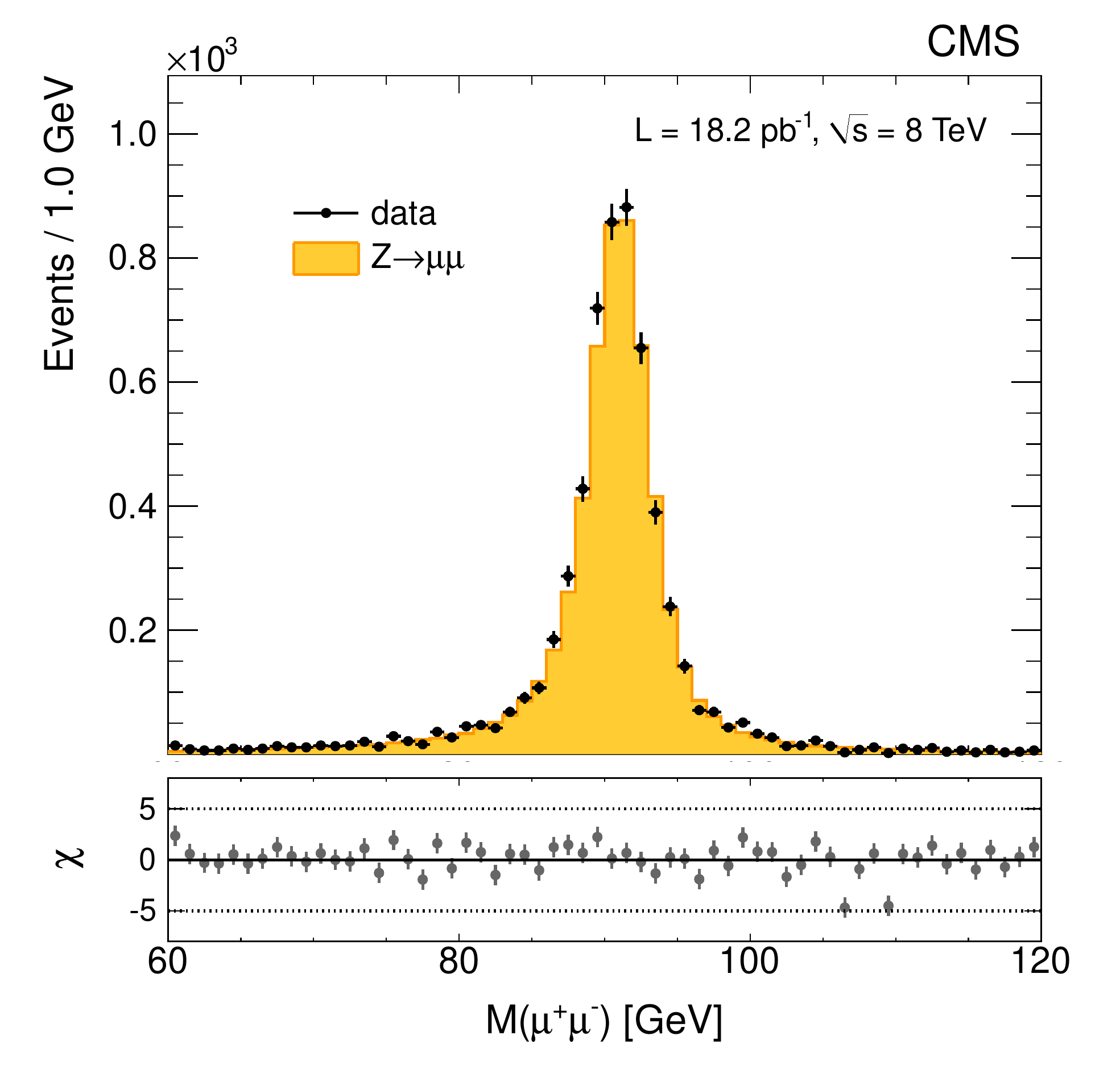}
       \caption{The dilepton mass distributions for $\cPZ$ boson candidate
       events in the electron (\cmsLeft) and muon (\cmsRight) final states. The variable $\chi$ shown in the
       lower plot is defined as
       $(N_{\text{obs}}-N_{\text{exp}})/\sqrt{N_{\text{obs}}}$,
       where $N_{\text{obs}}$ is the number of observed events and
       $N_{\text{exp}}$ is the total of the signal and background yields.
       \label{fig:Z}}
\end{figure}

To extract the $\cPZ$ boson yield, the events in the dilepton mass
window are counted. The yields contain a contribution of 3\% from
$\gamma^*$-mediated processes, including interference effects, as estimated with
\MCFM~\cite{Campbell:2010ff}. Background contamination is estimated from simulation to be about 0.4\%.
Figure~\ref{fig:Z} shows the dilepton mass distributions of the
inclusive $\cPZ$ samples. The signal yields, the acceptances, and the efficiencies are summarized in~\suppMaterial.
\nocite{UA1-wz, UA2-wz, CDF-wz-e, CDF-z-m, CDF-wz, D0-w} 

 The systematic uncertainties are summarized in Table~\ref{tab:systele} for the electron and muon channels.
The methods used to extract the systematic uncertainties for the
acceptance, efficiency, and signal extraction follow closely the $\PW$ and $\cPZ$ boson cross
section measurements performed at $\sqrt{s} =
7\TeV$~\cite{CMS-EWK-WZ}.
The leading experimental uncertainty comes from the measurement of the
 lepton reconstruction and identification efficiency. Other uncertainties come
 from the integrated luminosity of the data sample and theoretical
 uncertainties, which are dominated by the PDF uncertainties.

The luminosity of the data sample is measured with an uncertainty of 2.6\% by counting the number of
clusters per event in the silicon pixel detector. The highly granular detector,
consisting of ${\sim}$60 million channels, guarantees an excellent linearity of the
pixel detector response versus pileup. The method is calibrated by
means of a procedure pioneered by van~der~Meer~\cite{vanderMeer:296752}, consisting of beam scans along the
vertical and horizontal directions.
This van~der~Meer technique determines the luminosity at the percent
level from a measurement of the beam parameters~\cite{CMS-PAS-LUM-12-001}. The dominant contribution to the luminosity uncertainty originates from the assumptions on the functional form of the beam shapes.

\begin{table*}[tbh]
\topcaption{ \label{tab:systele}
Systematic uncertainties in percent for the electron and muon channels;
``\NA'' means that the source either does not apply or is negligible.}
\centering
\resizebox{\textwidth}{!}{
\begin {scotch} {lcccccccccccc}
 &
\multicolumn{2}{c}{$\PWp$} &
\multicolumn{2}{c}{$\PWm$} &
\multicolumn{2}{c}{$\PW$} &
\multicolumn{2}{c}{$\PWp$/$\PWm$} &
\multicolumn{2}{c}{$\cPZ$} &
\multicolumn{2}{c}{$\PW$/$\cPZ$} \\
Sources & $\Pe$ & $\mu$ & $\Pe$ & $\mu$ & $\Pe$ & $\mu$ & $\Pe$ & $\mu$ & $\Pe$ & $\mu$
& $\Pe$ & $\mu$ \\ \hline
Lepton reconstruction \& identification  & 2.8 & 1.0 & 2.5 & 0.9 & 2.5
& 1.0 & 3.8 & 1.2 &  2.8 & 1.1 & 3.8 & 1.5 \\
Momentum scale \& resolution             & 0.4 & 0.3 & 0.7 & 0.3 & 0.5
& 0.3 & 0.3 & 0.1 &  \NA & \NA & 0.5 & 0.3 \\
$\MET$ scale \& resolution               & 0.8 & 0.5 & 0.7 & 0.5 & 0.8
& 0.5 & 0.3 & 0.1 &  \NA & \NA & 0.8 & 0.5 \\
Background subtraction / modeling        & 0.2 & 0.2 & 0.3 & 0.1 & 0.3
& 0.1 & 0.1 & 0.2 &  0.4 & 0.4 & 0.5 & 0.4 \\
\hline
Total experimental                       & 3.0 & 1.2 & 2.7 & 1.1 & 2.7
& 1.2 & 3.8 & 1.2 &  2.8 & 1.2 & 3.9 & 1.7 \\
\hline
Theoretical uncertainty                  & 2.1 & 2.0 & 2.6 & 2.5 & 2.7
& 2.2 & 1.5 & 1.4 &  2.6 & 1.9 & 2.0 & 2.5 \\
\hline
Luminosity                               & 2.6 & 2.6 & 2.6 & 2.6 & 2.6
& 2.6 & \NA & \NA &  2.6 & 2.6 & \NA & \NA \\
\hline
Total                                    & 4.5 & 3.5 & 4.6 & 3.8 & 4.6
& 3.6 & 4.1 & 1.8 &  4.6 & 3.4 & 4.4 & 3.0 \\
\end{scotch}
}
\end{table*}

\begin{table*}[tbh]
\centering
\caption{ \label{tab:results}
Summary of total and fiducial $\PWp$, $\PWm$, $\PW$, and $\cPZ$ production cross sections times
branching fractions, $\PW$ to $\cPZ$ and $\PWp$ to $\PWm$ ratios, and their
theoretical predictions.}
\resizebox{\textwidth}{!}{
\begin {scotch} {llrllr}
\multicolumn{1}{c}{Channel} & \multicolumn{1}{c}{$\sigma \times \mathcal{B}$
[nb] (total)} & \multicolumn{1}{c}{NNLO [nb]} & \multicolumn{1}{c}{Quantity} & \multicolumn{1}{c}{Ratio (total)} & \multicolumn{1}{c}{NNLO} \\ \hline
$\PWp$ & $7.11 \pm 0.03\stat\pm 0.14\syst
\pm 0.18\lum$ & $7.12 \pm 0.20$ & $R_{\PWp/\PWm}$   & $1.39 \pm 0.01\stat\pm 0.02\syst$ & $1.41 \pm 0.01$ \\
$\PWm$ & $5.09 \pm 0.02\stat\pm 0.11\syst
\pm 0.13\lum$ & $5.06 \pm 0.13$ & $R_{\PW/\cPZ}$   & $10.63 \pm
0.11\stat\pm 0.25\syst$ & $10.74 \pm 0.04$ \\
$\PW$ & $12.21 \pm 0.03\stat\pm 0.24\syst \pm
0.32\lum$ & $12.18 \pm 0.32$ & & & \\
$\cPZ$ & $1.15 \pm 0.01\stat\pm 0.02\syst \pm 0.03\lum$ & $1.13 \pm
0.04$ & & & \\
\hline
\multicolumn{1}{c}{Channel} & \multicolumn{1}{c}{$\sigma \times
\mathcal{B}$ [nb] (fiducial)} & \multicolumn{1}{c}{NNLO [nb]} & \multicolumn{1}{c}{Quantity} & \multicolumn{1}{c}{Ratio (fiducial)} & \multicolumn{1}{c}{NNLO}\\ \hline
$\PWp$ & $3.16 \pm 0.01\stat\pm 0.04\syst
\pm 0.08\lum$ & $3.18 \pm 0.10$ & $R_{\PWp/\PWm}$ & $1.40 \pm 0.01\stat\pm 0.02\syst$ & $ 1.42 \pm 0.02$ \\
$\PWm$ & $2.26 \pm 0.01\stat\pm 0.02\syst
\pm 0.06\lum$ & $2.25 \pm 0.07$ & $R_{\PW/\cPZ}$    & $13.26 \pm 0.15\stat\pm 0.21\syst$ & $13.49 \pm 0.28$\\
$\PW$ & $5.42 \pm 0.02\stat\pm 0.06\syst \pm 0.14\lum$ & $5.43 \pm 0.16$\\
$\cPZ$ & $0.41 \pm 0.01\stat\pm 0.01\syst \pm 0.01\lum$ & $0.40 \pm 0.01$ \\
\end{scotch}
}
\end{table*}

The theoretical predictions of cross sections and cross section ratios are computed at NNLO with the program
\textsc{fewz}~\cite{FEWZ} and the MSTW2008~\cite{Martin:2009iq} set of PDFs. The
uncertainties in these predictions, at the 68\% confidence level (\CL), include
contributions from the uncertainty of the strong coupling constant $\alpha_s$~\cite{PDF_alphaS1, PDF_alphaS2}, the choice of heavy-quark masses (charm and bottom quarks)~\cite{PDF_heavy_quarks}, as well as neglected higher-order corrections
beyond NNLO, which are estimated by allowing the renormalization and factorization scales
to vary. The NNLO predictions for the total cross sections times branching fractions are $\THEORYSIGBRWP$ for $\PWp$, $\THEORYSIGBRWM$ for $\PWm$, and
$\THEORYSIGBRZ$ for $\cPZ$ boson production. The $\cPZ$ boson cross section
requires an invariant mass within the range 60 to 120\GeV, and it
includes the effects of virtual photons.

\begin{figure*}[tbh]
\centering
  \includegraphics[width=0.48\textwidth]{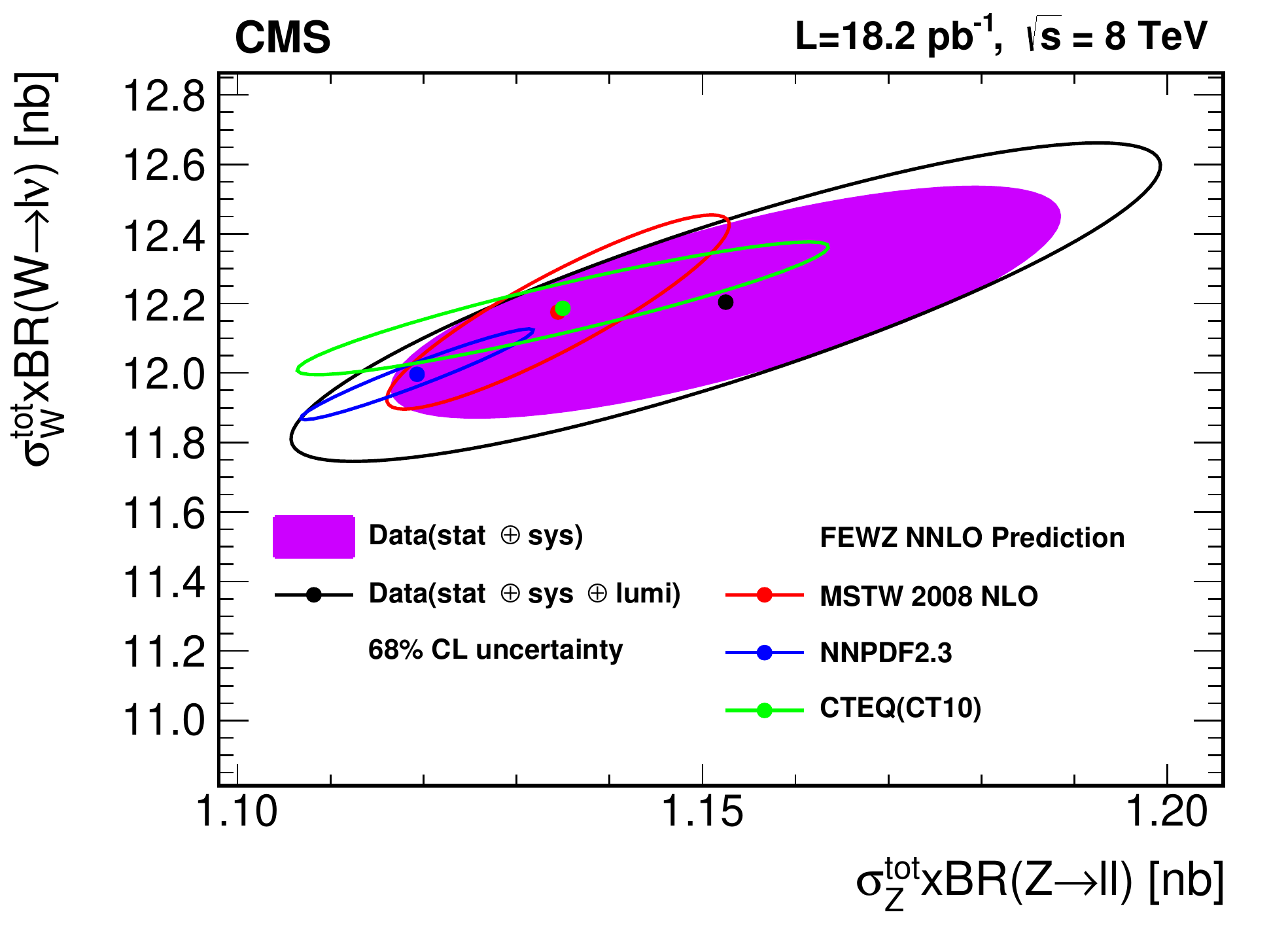}
  \includegraphics[width=0.48\textwidth]{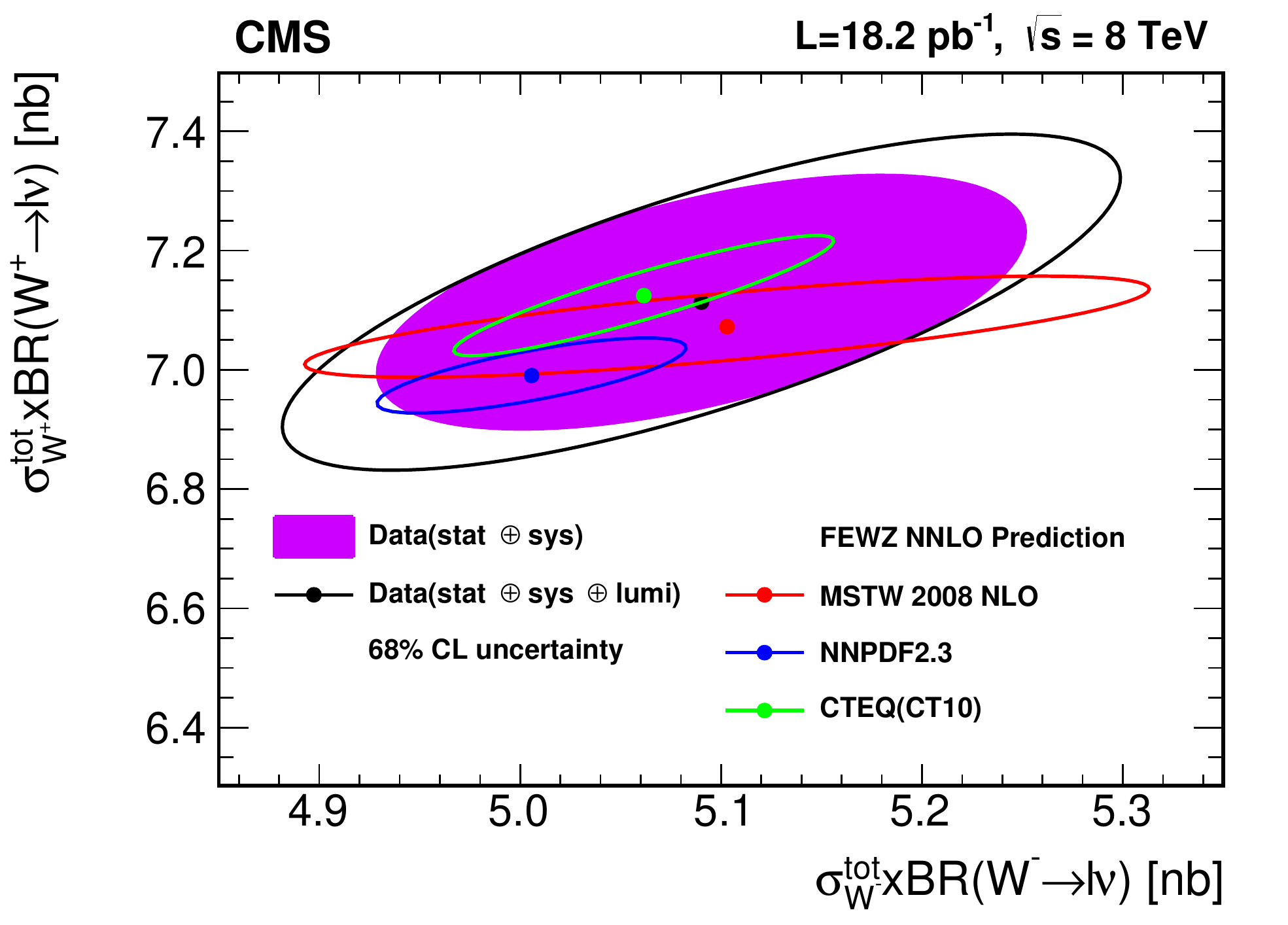}
  \includegraphics[width=0.48\textwidth]{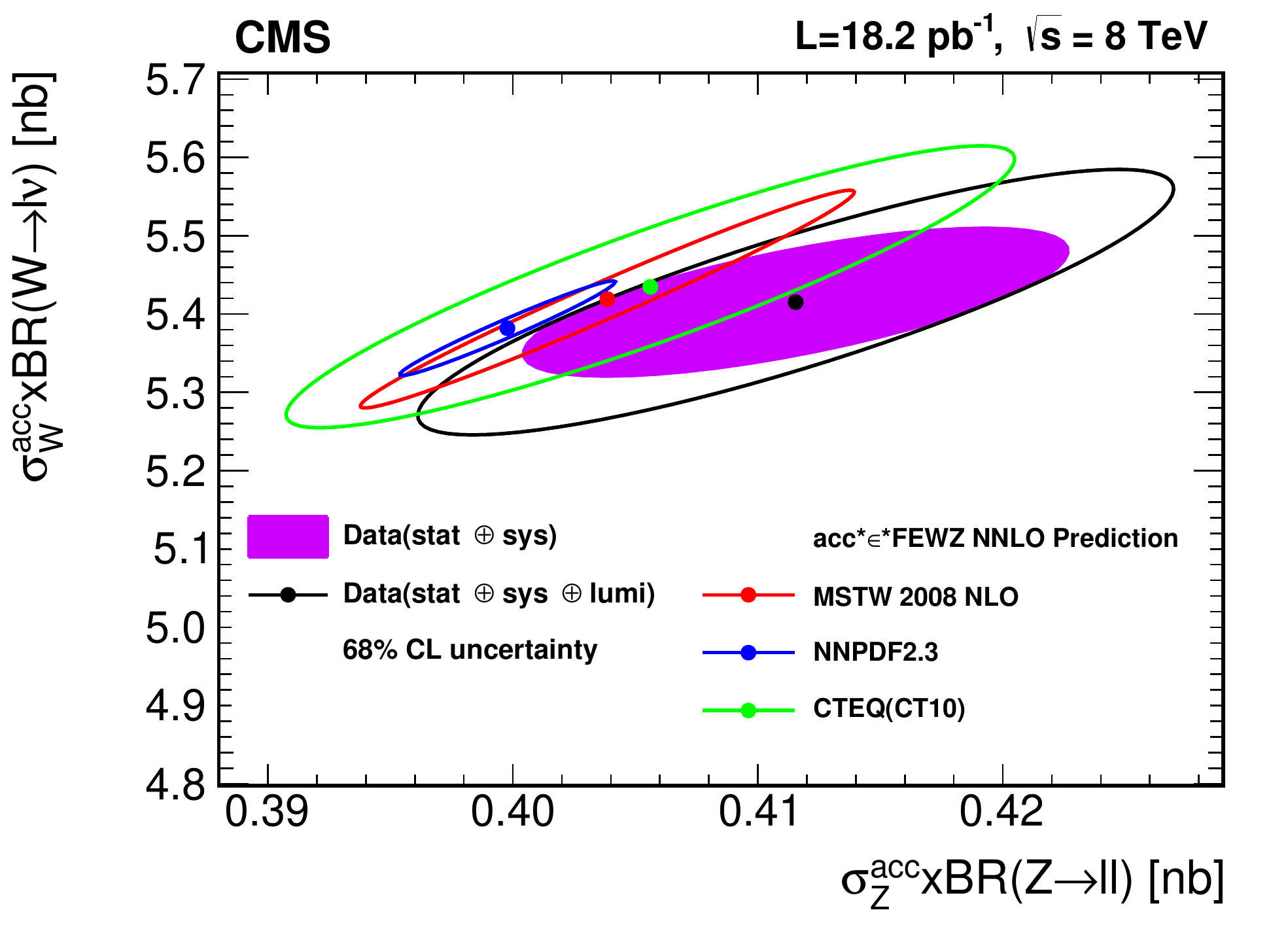}
  \includegraphics[width=0.48\textwidth]{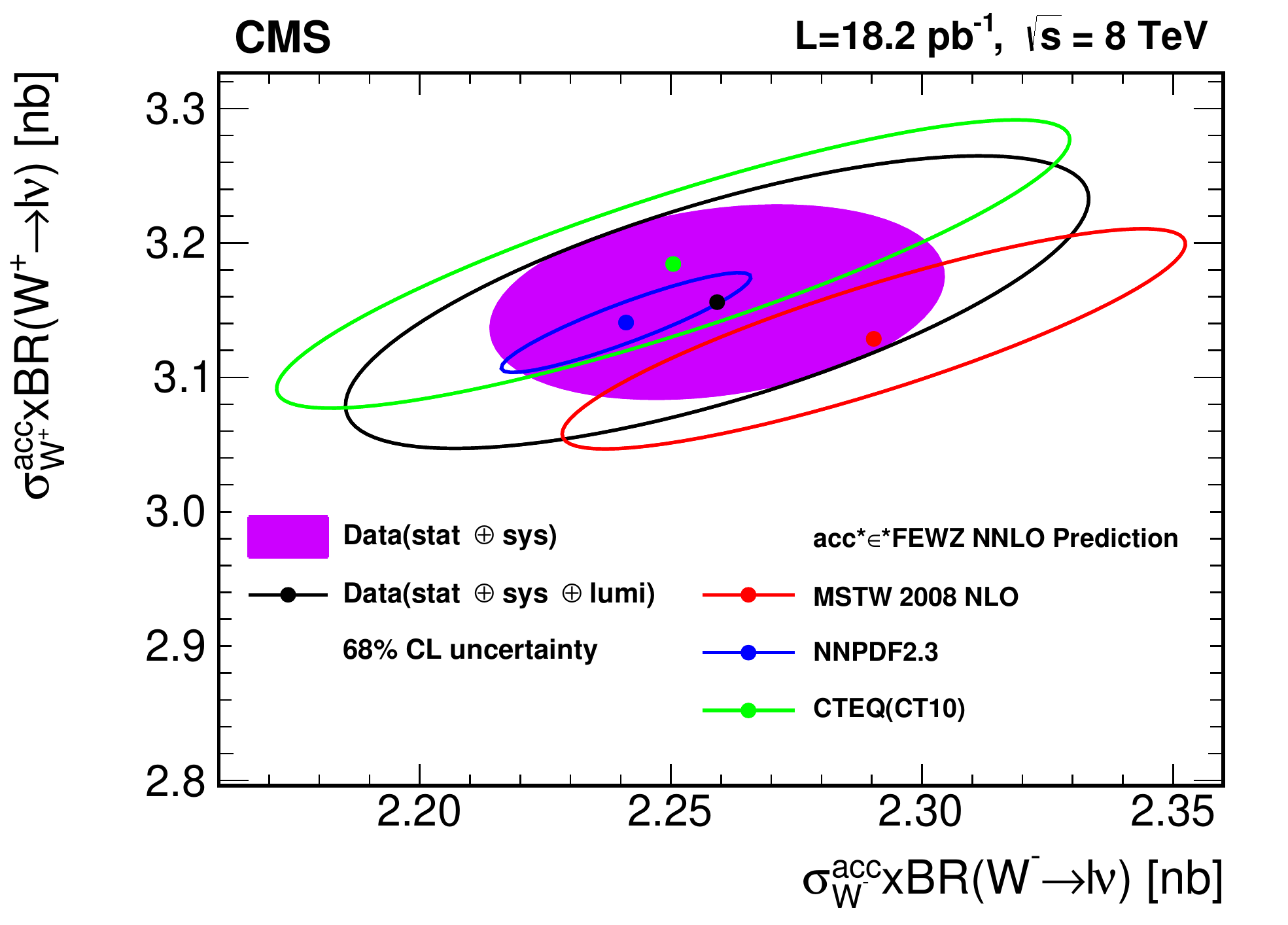}
 \caption{\label{fig:PDF_WZratio}
Measured and predicted $\PW$ versus $\cPZ$ boson (left column) and
$\PWp$ versus $\PWm$ boson (right column) production cross sections
times branching fractions. The ellipses illustrate the 68\% \CL coverage for total uncertainties (open) and excluding the luminosity uncertainty (filled). The
top row shows the inclusive cross sections times branching fractions and
the bottom row shows the results within the fiducial regions. The uncertainties in the theoretical predictions correspond to the PDF
uncertainty components only and are evaluated for MSTW 2008 NLO~\cite{Martin:2009iq}, NNPDF 2.3~\cite{NNPDF23}, and CT10~\cite{Lai:2010vv}.}
\end{figure*}

The results in the electron and muon decay channels are compatible with a p-value of 0.42. Assuming universality of lepton couplings to $\PW$ and $\cPZ$ bosons, the channels are combined by calculating an average cross section value weighted by their statistical and systematic uncertainties, taking into account the correlated uncertainties. The two leptonic decay channels
are combined by assuming fully correlated uncertainties for the
acceptance and luminosity, but with other uncertainties assumed to be uncorrelated.

In measurements of the ratios of cross sections some systematic
uncertainties cancel, most importantly the uncertainty in the
luminosity. The uncertainties in the lepton reconstruction and
identification are treated as uncorrelated and the resulting experimental
uncertainty in the ratio measurements can therefore be larger than for individual cross section measurements.
A summary of the measurements is given in Table~\ref{tab:results},
including the results obtained within the fiducial regions in \pt and
$\eta$. See~\suppMaterial for the total cross sections times branching fractions and ratios
for the electron and muon decay channels.

The upper two plots in Fig.~\ref{fig:PDF_WZratio} show the measured and predicted $\PW$ versus
$\cPZ$ and $\PWp$ versus $\PWm$ cross sections for different PDF sets.
The uncertainties in the theoretical predictions correspond to the PDF
uncertainties only. This approach eliminates the need to propagate acceptance uncertainties originating from the PDFs and higher-order corrections into the measurement.
The final measurement is compared with the predicted cross sections in
the acceptance region for three different PDFs with their respective uncertainty bands propagated through the prediction.

In summary, we have performed the first measurements of total and
fiducial inclusive $\PW$ and $\cPZ$ production cross
sections times branching fractions in $\Pp\Pp$ collisions at $\sqrt{s} = 8\TeV$
using $\THELUMI$ of low-pileup data recorded with the CMS detector. The $\PW$ and $\cPZ$ bosons are observed via their decays to
electrons and muons. The measured total inclusive production cross sections times branching fractions are
$\sigma(\Pp\Pp \to \PW X) \times \mathcal{B}(\PW \to \ell\nu ) =
12.21 \pm 0.03\stat\pm 0.24\syst\pm
0.32\lum\unit{nb}$ and, for the dilepton mass in the range of 60
to 120\GeV, $\sigma(\Pp\Pp \to \cPZ X) \times \mathcal{B}(\cPZ \to \ell^+\ell^-) = 1.15 \pm 0.01\stat\pm 0.02\syst \pm 0.03\lum\unit{nb}$.
In addition to the inclusive
cross sections, we present ratios of cross sections measured with a
precision of 2\%.
The measurements in the electron and muon channels
are consistent, and in agreement with NNLO calculations. Additional figures summarizing our measurements are available in~\suppMaterial.

We congratulate our colleagues in the CERN accelerator departments for the excellent performance of the LHC and thank the technical and administrative staffs at CERN and at other CMS institutes for their contributions to the success of the CMS effort. In addition, we gratefully acknowledge the computing centres and personnel of the Worldwide LHC Computing Grid for delivering so effectively the computing infrastructure essential to our analyses. Finally, we acknowledge the enduring support for the construction and operation of the LHC and the CMS detector provided by the following funding agencies: BMWF and FWF (Austria); FNRS and FWO (Belgium); CNPq, CAPES, FAPERJ, and FAPESP (Brazil); MES (Bulgaria); CERN; CAS, MoST, and NSFC (China); COLCIENCIAS (Colombia); MSES and CSF (Croatia); RPF (Cyprus); MoER, SF0690030s09 and ERDF (Estonia); Academy of Finland, MEC, and HIP (Finland); CEA and CNRS/IN2P3 (France); BMBF, DFG, and HGF (Germany); GSRT (Greece); OTKA and NIH (Hungary); DAE and DST (India); IPM (Iran); SFI (Ireland); INFN (Italy); NRF and WCU (Republic of Korea); LAS (Lithuania); MOE and UM (Malaysia); CINVESTAV, CONACYT, SEP, and UASLP-FAI (Mexico); MBIE (New Zealand); PAEC (Pakistan); MSHE and NSC (Poland); FCT (Portugal); JINR (Dubna); MON, RosAtom, RAS and RFBR (Russia); MESTD (Serbia); SEIDI and CPAN (Spain); Swiss Funding Agencies (Switzerland); NSC (Taipei); ThEPCenter, IPST, STAR and NSTDA (Thailand); TUBITAK and TAEK (Turkey); NASU (Ukraine); STFC (United Kingdom); DOE and NSF (USA).

\bibliography{auto_generated}   
\ifthenelse{\boolean{cms@external}}
{}
{
\clearpage
\appendix
\section[Supplemental material]{Total cross sections times branching fractions and ratios
for the electron and muon decay channels}
\newcommand\PWEMYIELD{30857 \pm 185 }
\newcommand\PWEPYIELD{44194 \pm 219 }
\newcommand\PWEIYIELD{75051 \pm 287 }
\newcommand\PWMMYIELD{33836 \pm 182 }
\newcommand\PWMPYIELD{47637 \pm 216 }
\newcommand\PWMIYIELD{81473 \pm 282 }
\label{sec:SuppMaterial}
\begin{table*}[thbp]
\topcaption{ \label{tab:WesignalCollection}
The signal yields, acceptances, and
efficiencies. The $\cPZ$ boson yield uncertainties are
given by Poisson statistics, while the $\PW$ boson yield uncertainties are
determined from the fit. Uncertainties in the acceptances and
efficiencies are systematic uncertainties.}
\centering
\begin {scotch} {lcccc}
Source     &  $\cPZ\to\Pep\Pem$  & $\PW\to\Pe\Pgn$           & $\PWp\to\Pep\Pgn$           & $\PWm\to\Pem\Pagn$            \\
\hline
Yields     & $4793 \pm 69$ & $\PWEIYIELD$      &  $\PWEPYIELD$      & $\PWEMYIELD$      \\
Acceptance & 0.399 $\pm$ 0.010 & 0.479 $\pm$ 0.013 & 0.484 $\pm$ 0.011 & 0.471 $\pm$ 0.013 \\
Efficiency & 0.585 $\pm$ 0.016 & 0.695 $\pm$ 0.019 & 0.687 $\pm$ 0.021 & 0.708 $\pm$ 0.019\\
 \hline
Source      &  $\cPZ\to\Pgmp\Pgmm$  & $\PW\to\Pgm\Pgn$           & $\PWp\to\Pgmp\Pgn$           & $\PWm\to\Pgmm\Pagn$            \\
\hline
Yields    & 5917 $\pm$ 77 & $\PWMIYIELD$      &  $\PWMPYIELD$      &  $\PWMMYIELD$      \\
Acceptance & 0.346 $\pm$ 0.007 & 0.440 $\pm$ 0.010 & 0.441 $\pm$ 0.009 & 0.439 $\pm$ 0.011\\
Efficiency & 0.809 $\pm$ 0.010 & 0.839 $\pm$ 0.010 & 0.843 $\pm$ 0.010 & 0.831 $\pm$ 0.009\\
\end{scotch}
\end{table*}

\begin{table*}[tbhp]
\centering
\topcaption{ \label{tab:resultsApp}
Summary of total $\PWp$, $\PWm$, $\PW$, and $\cPZ$ production cross sections times
branching fractions, $\PW$ to $\cPZ$ and $\PWp$ to $\PWm$ ratios, and their
theoretical predictions.}
\begin {scotch} {lllr}
\multicolumn{2}{c}{Channel} & \multicolumn{1}{c}{$\sigma \times \mathcal{B}$
[nb] (total)} & \multicolumn{1}{c}{NNLO [nb]} \\
\hline
      & $\Pep\nu$ & $7.31 \pm 0.04\stat\pm 0.26\syst \pm 0.19\lum$ & \\
$\PWp$ & $\mu^+\nu$ & $7.04 \pm 0.03\stat\pm 0.16\syst \pm 0.18\lum$ & $7.12 \pm 0.20$\\
      & $\ell^+\nu$ & $7.11 \pm 0.03\stat\pm 0.14\syst \pm 0.18\lum$ & \\\hline
      & $\Pem\nu$ & $5.08 \pm 0.03\stat\pm 0.19\syst \pm 0.13\lum$ & \\
$\PWm$ & $\mu^-\nu$ & $5.09 \pm 0.03\stat\pm 0.14\syst \pm 0.13\lum$ & $5.06 \pm 0.13$\\
      & $\ell^-\nu$ & $5.09 \pm 0.02\stat\pm 0.11\syst \pm 0.13\lum$ & \\\hline
      & $\Pe\nu$ & $12.39 \pm 0.05\stat\pm 0.44\syst \pm 0.32\lum$ & \\
$\PW$  & $\mu\nu$ & $12.13 \pm 0.04\stat\pm 0.29\syst \pm 0.32\lum$ &  $12.18 \pm 0.32$ \\
      & $\ell\nu$ & $12.21 \pm 0.03\stat\pm 0.24\syst \pm 0.32\lum$ & \\\hline
    & $\Pep\Pem$ & $1.13 \pm 0.02\stat\pm 0.05\syst \pm 0.03\lum$ & \\
$\cPZ$& $\mu^+\mu^-$ & $1.16 \pm 0.02\stat\pm 0.03\syst \pm 0.03\lum$ & $1.13 \pm 0.04$\\
    & $\ell^+\ell^-$& $1.15 \pm 0.01\stat\pm 0.02\syst \pm 0.03\lum$ & \\\hline
\multicolumn{2}{c}{Quantity} & \multicolumn{1}{c}{Ratio (total)} & \multicolumn{1}{c}{NNLO} \\ \hline
& $\Pe$ & $1.44 \pm 0.01\stat\pm 0.05\syst$ & \\
$R_{\PWp/\PWm}$ & $\mu$ & $1.38 \pm 0.01\stat\pm 0.03\syst$ & $1.41 \pm 0.01$ \\
  & $\ell$ & $1.39 \pm 0.01\stat\pm 0.02\syst$ & \\
\hline
             & $\Pe$   & $10.99 \pm 0.16\stat\pm 0.43\syst$ &                  \\
$R_{\PW/\cPZ}$   & $\mu$ & $10.44 \pm 0.14\stat\pm 0.30\syst$ & $10.74 \pm 0.04$ \\
             & $\ell$ & $10.63 \pm 0.11\stat\pm 0.25\syst$ &  \\
\end{scotch}
\end{table*}

\begin{table*}[htbp]
\topcaption{ \label{tab:resultsAppFid}
Summary of fiducial inclusive $\PWp$, $\PWm$, $\PW$, and $\cPZ$ production cross sections times
branching fractions, $\PW$ to $\cPZ$ and $\PWp$ to $\PWm$ ratios, and their
theoretical predictions.}
\centering
\begin {scotch} {lllr}
\multicolumn{2}{c}{Channel} & \multicolumn{1}{c}{$\sigma \times \mathcal{B}$
[nb] (total)} & \multicolumn{1}{c}{NNLO [nb]} \\
\hline
      & $\Pep\nu$ & $3.54 \pm 0.02\stat\pm 0.11\syst\pm 0.09\lum$ & $3.45 \pm 0.12$\\
$\PWp$ & $\mu^+\nu$ & $3.10 \pm 0.01\stat\pm 0.04\syst \pm 0.08\lum$ & $3.14 \pm 0.11$\\
      & $\ell^+\nu$ & $3.16 \pm 0.01\stat\pm 0.04\syst\pm 0.08\lum$ & $3.18 \pm 0.10$ \\
\hline
      & $\Pem\nu$ & $2.39 \pm 0.01\stat\pm 0.06\syst \pm 0.06\lum$ & $2.38 \pm 0.09$ \\
$\PWm$ & $\mu^-\nu$ & $2.24 \pm 0.01\stat\pm 0.02\syst \pm 0.06\lum$ & $2.22 \pm 0.08$\\
      & $\ell^-\nu$ & $2.26 \pm 0.01\stat\pm 0.02\syst \pm 0.06\lum$ & $2.25 \pm 0.07$\\\hline
      & $\Pe\nu$ & $5.94 \pm 0.02\stat\pm 0.16\syst \pm 0.15\lum$ & $5.84 \pm 0.22$\\
$\PW$  & $\mu\nu$ & $5.34 \pm 0.02\stat\pm 0.06\syst \pm 0.14\lum$ & $5.36 \pm 0.18$ \\
      & $\ell\nu$ & $5.42 \pm 0.02\stat\pm 0.06\syst \pm 0.14\lum$ & $5.43 \pm 0.16$\\\hline
    & $\Pep\Pem$ & $0.45 \pm 0.01\stat\pm 0.01\syst \pm 0.01\lum$ & $0.45 \pm 0.02$\\
$\cPZ$& $\mu^+\mu^-$ & $0.40 \pm 0.01\stat\pm 0.01\syst \pm 0.01\lum$ & $0.39 \pm 0.02$\\
    & $\ell^+\ell^-$& $0.41 \pm 0.01\stat\pm 0.01\syst \pm 0.01\lum$ & $0.40 \pm 0.01$\\\hline
\multicolumn{2}{c}{Quantity} & \multicolumn{1}{c}{Ratio (total)} & \multicolumn{1}{c}{NNLO} \\ \hline
& $\Pe$ & $1.48 \pm 0.01\stat\pm 0.06\syst$ & $1.45 \pm 0.02$ \\
$R_{\PWp/\PWm}$ & $\mu$ & $1.39 \pm 0.01\stat\pm 0.02\syst$ & $1.41 \pm 0.02$ \\
  & $\ell$ & $1.40 \pm 0.01\stat\pm 0.02\syst$ & $1.42 \pm 0.02$ \\ \hline
             & $\Pe$   & $13.19 \pm 0.20\stat\pm 0.51\syst$ & $12.89 \pm 0.26$ \\
$R_{\PW/\cPZ}$   & $\mu$ & $13.28 \pm 0.18\stat\pm 0.23\syst$ & $13.65 \pm 0.35$ \\
             & $\ell$ & $13.26 \pm 0.15\stat\pm 0.21\syst$ &  $13.49 \pm 0.28$ \\
\end{scotch}
\end{table*}

\begin{figure}[tbh]
\centering
\includegraphics[width=.85\columnwidth]{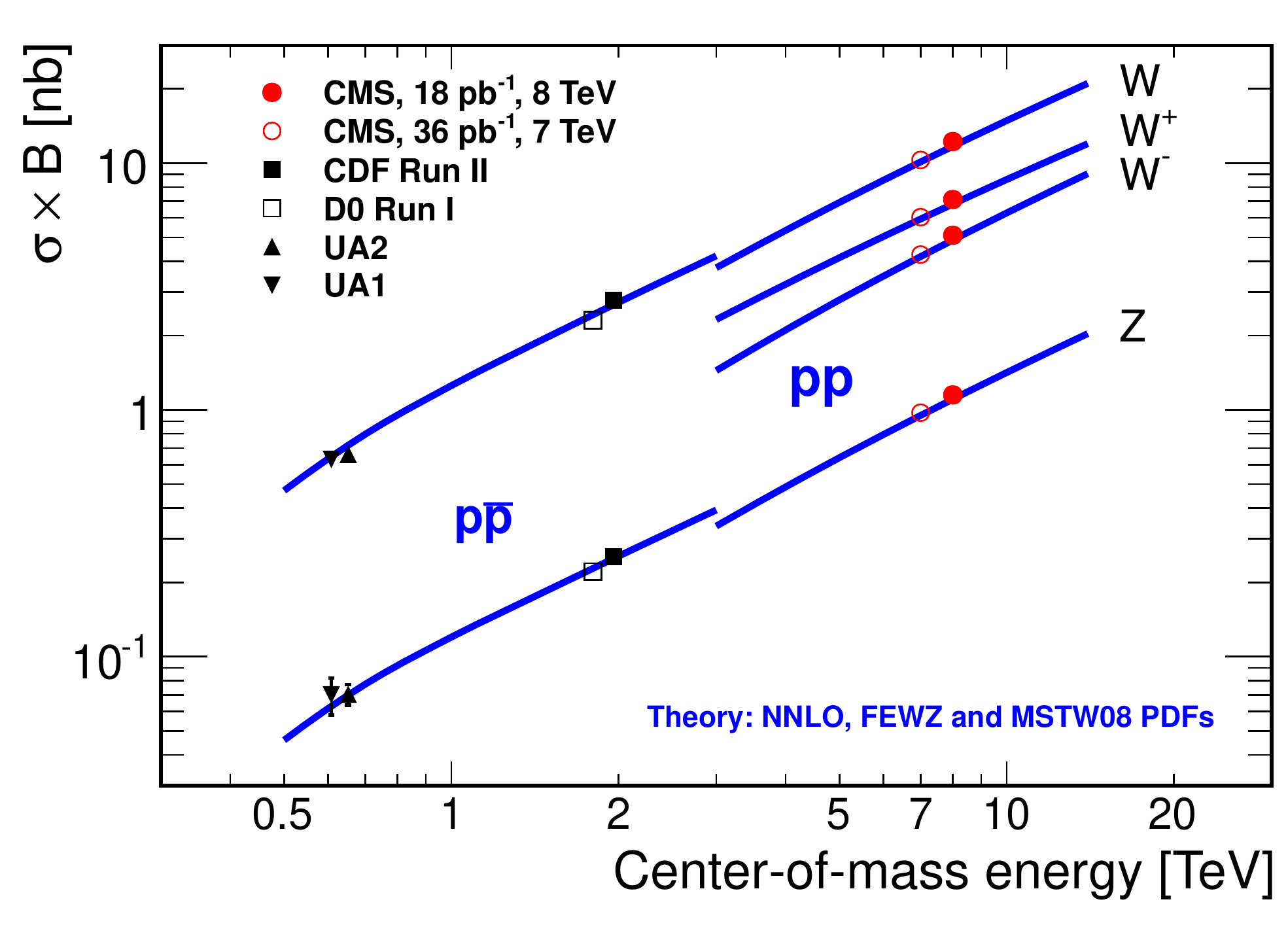}
\caption{Measurements of the total $\PWp$, $\PWm$, $\PW$, and $\cPZ$ production cross
       sections times branching fractions versus center-of-mass energy for CMS and experiments at lower-energy
       colliders~\cite{UA1-wz, UA2-wz, CDF-wz-e, CDF-z-m, CDF-wz,
         D0-w}. The predicted increase of the cross sections with center-of-mass energy is confirmed by our measurements. 
       \label{fig:collidersApp}}
\end{figure}

\begin{figure}[htbp]
\centering
\includegraphics[width=.85\columnwidth]{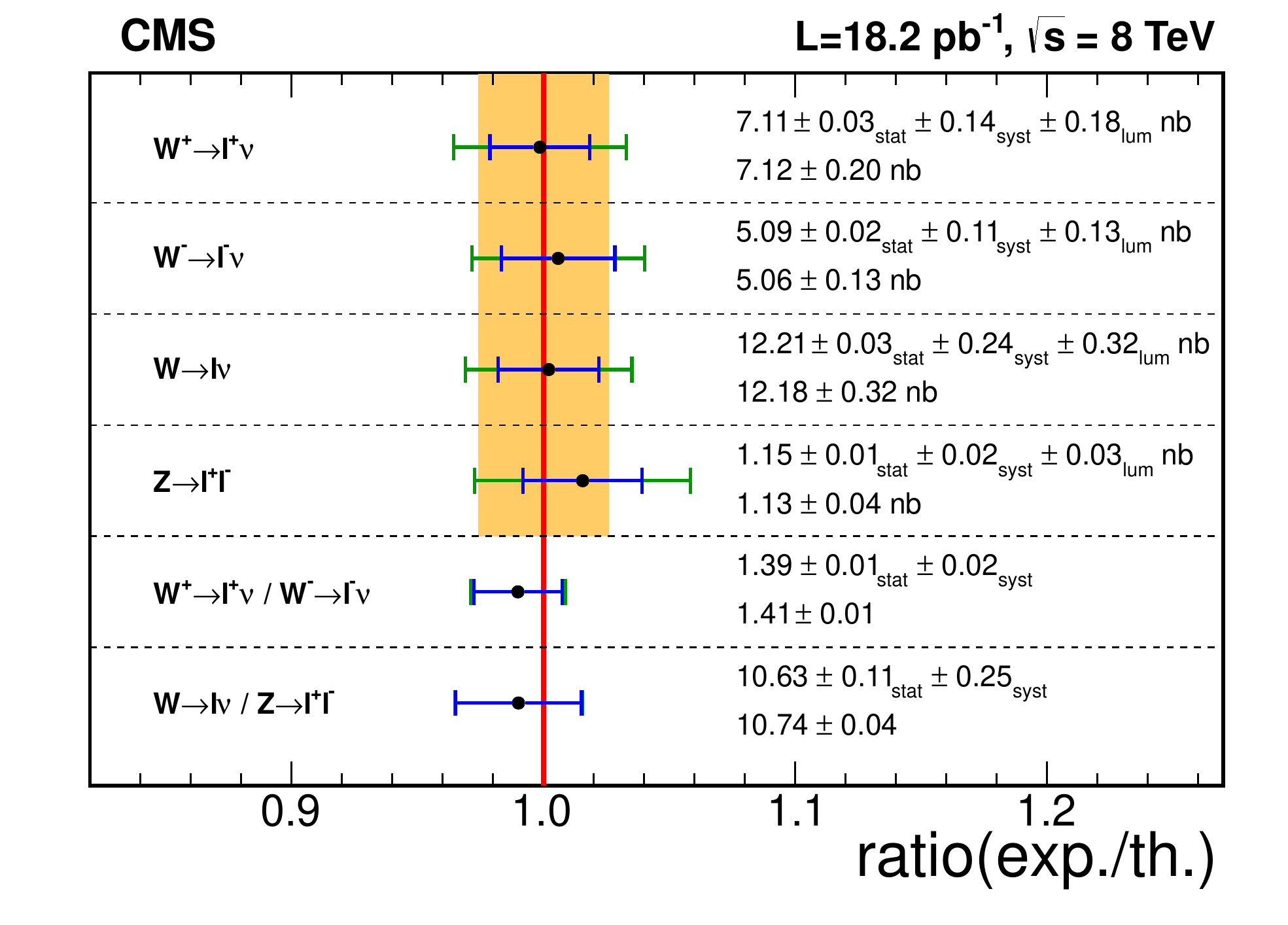}
\caption{Summary of total inclusive $\PWp$, $\PWm$, $\PW$, and $\cPZ$ production cross sections times
branching fractions, $\PW$ to $\cPZ$ and $\PWp$ to $\PWm$ ratios, and their
theoretical predictions. The shaded box indicates
the uncertainties in the luminosity measurement. The inner error bars represent the experimental
uncertainties, while outer error bars also include the uncertainties in the theoretical predictions.
The individual measurements and theoretical predictions are given numerically on the
right.
\label{fig:resultsAppOne}}
\end{figure}

\begin{figure}[htbp]
\centering
\includegraphics[width=.85\columnwidth]{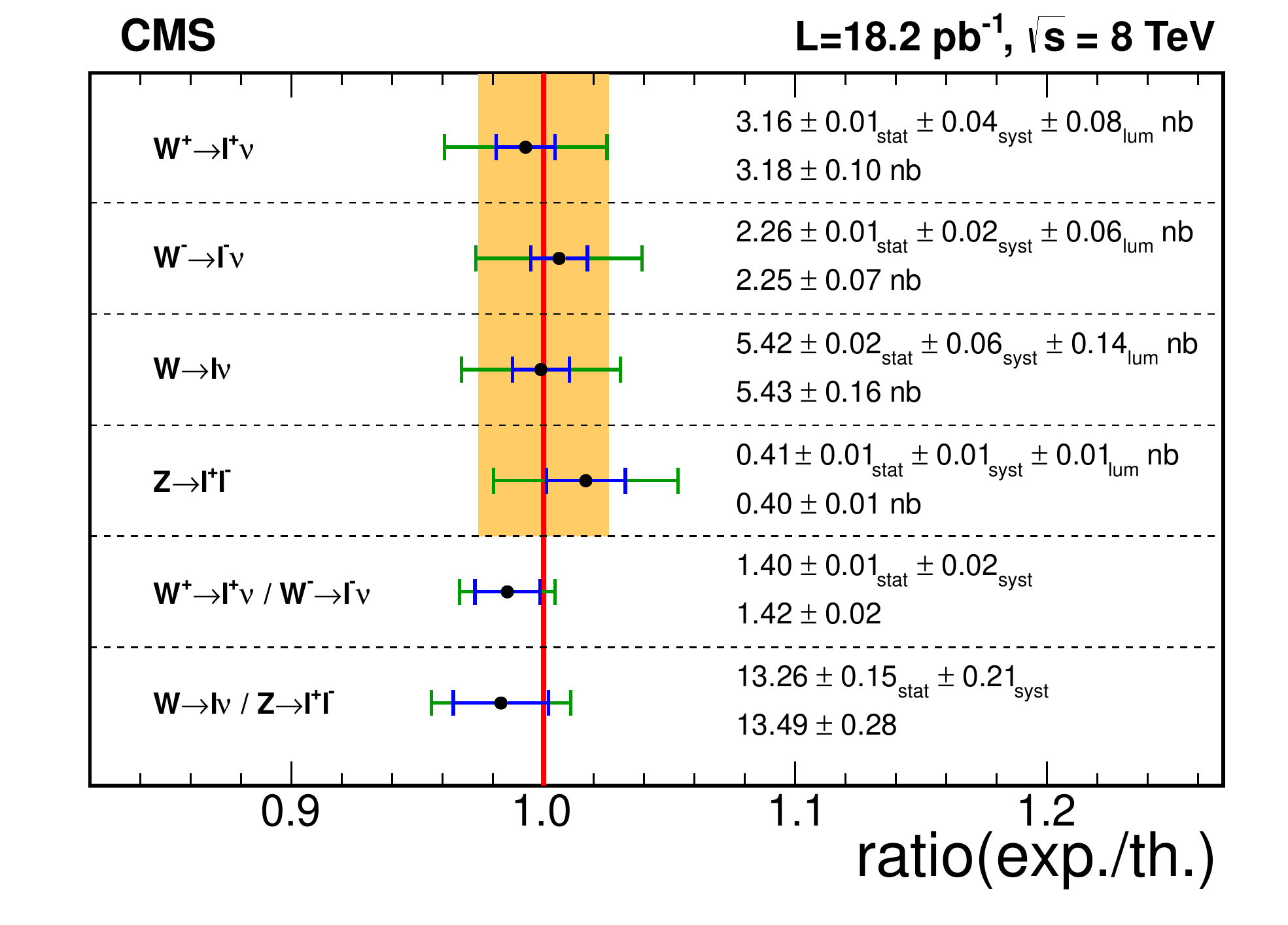}
\caption{Summary of fiducial inclusive $\PWp$, $\PWm$, $\PW$, and $\cPZ$ production cross sections times
branching fractions, $\PW$ to $\cPZ$ and $\PWp$ to $\PWm$ ratios, and their
theoretical predictions. The shaded box indicates
the uncertainties in the luminosity measurement. The inner error bars represent the experimental
uncertainties, while outer error bars also include the uncertainties in the theoretical predictions.
The individual measurements and theoretical predictions are given numerically on the
right. \label{fig:resultsAppTwo}}
\end{figure}
}
\cleardoublepage \section{The CMS Collaboration \label{app:collab}}\begin{sloppypar}\hyphenpenalty=5000\widowpenalty=500\clubpenalty=5000\textbf{Yerevan Physics Institute,  Yerevan,  Armenia}\\*[0pt]
S.~Chatrchyan, V.~Khachatryan, A.M.~Sirunyan, A.~Tumasyan
\vskip\cmsinstskip
\textbf{Institut f\"{u}r Hochenergiephysik der OeAW,  Wien,  Austria}\\*[0pt]
W.~Adam, T.~Bergauer, M.~Dragicevic, J.~Er\"{o}, C.~Fabjan\cmsAuthorMark{1}, M.~Friedl, R.~Fr\"{u}hwirth\cmsAuthorMark{1}, V.M.~Ghete, N.~H\"{o}rmann, J.~Hrubec, M.~Jeitler\cmsAuthorMark{1}, W.~Kiesenhofer, V.~Kn\"{u}nz, M.~Krammer\cmsAuthorMark{1}, I.~Kr\"{a}tschmer, D.~Liko, I.~Mikulec, D.~Rabady\cmsAuthorMark{2}, B.~Rahbaran, C.~Rohringer, H.~Rohringer, R.~Sch\"{o}fbeck, J.~Strauss, A.~Taurok, W.~Treberer-Treberspurg, W.~Waltenberger, C.-E.~Wulz\cmsAuthorMark{1}
% ...
\vskip\cmsinstskip
% ...
1:~~Also at Vienna University of Technology, Vienna, Austria\\
2:~~Also at CERN, European Organization for Nuclear Research, Geneva, Switzerland\\
% ...

\end{sloppypar}
\end{document}